\documentclass{PoS}

\usepackage{color}
\usepackage{amsthm}
\usepackage{mathtools}
\usepackage[normalem]{ulem} 
\usepackage{graphicx}
\usepackage{amssymb,latexsym}
\usepackage{amsmath}
\usepackage{amsfonts}
\usepackage{mathrsfs}
\usepackage{bbm}
\usepackage{bm}
\usepackage[T1]{fontenc}

\def\slasha#1{\setbox0=\hbox{$#1$}#1\hskip-\wd0\hbox to\wd0{\hss\sl/\/\hss}}

\def\periodb#1{\setbox0=\hbox{$#1$}#1\hskip-\wd0\hbox to\wd0{-}}

\newcommand{\CA}{\mathcal{A}}    			

\newcommand{\CF}{\mathcal{F}}

\newcommand{\CP}{\mathcal{P}}

\newcommand{\CV}{\mathcal{V}}

\newcommand{\CX}{\mathcal{X}}

\newcommand{\frg}{\mathfrak{g}}				

\newcommand{\FR}{\mathbbm{R}}     			
\newcommand{\RZ}{\mathbbm{Z}}     			

\newcommand{\dd}{\mathrm{d}}     			

\newcommand{\LL}{\mathrm{L}}     			

\newcommand{\sSO}{\mathsf{SO}}

\newcommand{\comment}[1]{}     				
\def\tyng(#1){\hbox{\tiny$\yng(#1)$}}			
\def\tyoung(#1){\hbox{\tiny$\young(#1)$}}			

\newcommand{\beq}{\begin{eqnarray}}
\newcommand{\eeq}{\end{eqnarray}}

\newcommand{\sff}{{\sf f}}

\newcommand{\sfR}{\mathsf{R}}

\definecolor{outrageousorange}{rgb}{1.0, 0.43, 0.29}

\newenvironment{myitemize}{
\vspace{-2mm}\begin{itemize}
\setlength{\itemsep}{-1mm}
}{\vspace{-2mm}\end{itemize}}

\newcommand{\Tr}{\mathrm{Tr}}

\theoremstyle{plain}

\theoremstyle{definition}

\theoremstyle{remark}

\newcommand{\dwedge}{\curlywedge}
\newcommand{\om}{\omega}
\newcommand{\epsi}{\epsilon}
\newcommand{\nn}{\nonumber}

\newcommand{\midwedge}{\text{\Large$\wedge$}}

\def\RR{{\mathcal R}}

\title{Homotopy Lie Algebras of Gravity \\ and their Braided Deformations}

\ShortTitle{Homotopy Lie Algebras of Gravity and their Braided
  Deformations}

\author{Marija Dimitrijevi\'c \'Ciri\'c\\
Faculty of Physics, University of Belgrade, Beograd, Serbia\\
E-mail: \email{dmarija@ipb.ac.rs}}

\author{\speaker{Grigorios Giotopoulos}\\
Department of Mathematics, Heriot-Watt University, Edinburgh, United
Kingdom\\
Maxwell Institute for
	Mathematical Sciences, Edinburgh, United Kingdom\\
Email: \email{gg42@hw.ac.uk}}

\author{Voja Radovanovi\'c\\
Faculty of Physics, University of Belgrade, Beograd, Serbia\\
E-mail: \email{rvoja@ipb.ac.rs}}

\author{Richard J. Szabo\\
Department of Mathematics, Heriot-Watt University, Edinburgh, United
Kingdom\\
Maxwell Institute for
	Mathematical Sciences, Edinburgh, United Kingdom\\
Higgs Centre
	for Theoretical Physics, Edinburgh, United Kingdom\\
Email: \email{R.J.Szabo@hw.ac.uk}}

\abstract{We describe the cyclic $L_{\infty}$-algebra
  formulation of classical general relativity without matter fields in the
  Einstein-Cartan-Palatini formalism. Using Drinfel'd twist
  deformation techniques, we define a noncommutative version of the theory with
  braided gauge symmetries. We introduce a notion of braided
  $L_{\infty}$-algebra, and use it to encode the 
  symmetries, field content, field equations and Noether identities of
  noncommutative gravity.}

\FullConference{Corfu Summer Institute 2019 "School and Workshops on Elementary Particle Physics and Gravity" (CORFU2019)\\
		31 August -- 25 September 2019\\
		Corf\`u, Greece\\
Preprint: \  {\tt EMPG--20--10}}

\begin{document}

\section{Introduction}

Recent advances in string theory suggest that the
low-energy effective dynamics of closed strings in non-geometric flux
compactifications may be described by noncommutative and even
nonassociative deformations of
gravity~\cite{Blumenhagen:2010hj,Lust:2010iy,Blumenhagen:2011ph,Mylonas:2012pg,Blumenhagen:2013zpa}. The
framework of nonassociative differential geometry to accommodate such situations was developed
in~\cite{Mylonas:2013jha,Aschieri:2015roa,Blumenhagen:2016vpb,Aschieri:2017sug}. However,
a suitable generalization of the
Einstein-Hilbert action in the traditional metric formulation of
general relativity has not yet been found. On the other hand,
noncommutative and nonassociative deformations of gravity can be
studied within the first
order formulation of general relativity, which treats the theory as a
deformation of a gauge theory and in which a deformed action can be
constructed~\cite{Barnes:2015uxa,Barnes:2016cjm}.
However, the deformed symmetries of these models do not behave as
conventional gauge symmetries and have not been thorougly understood: an easy calculation shows that, with
the naive definition of gauge transformations, nonassociativity
obstructs closure of the gauge algebra. 

The lack of closure of gauge transformations naturally suggests an
approach based on $L_\infty$-algebras, which have a long history in
both physics and mathematics. The main motivation behind this
contribution, which is based on a series of papers~\cite{ECPLinfty,Inprep}, is to understand the
symmetries, field equations and their Lagrangian formulations for
noncommutative and nonassociative deformations of gravity via
the formalism of $L_{\infty}$-algebras. $L_\infty$-algebras are generalizations of differential graded Lie algebras with infinitely-many graded antisymmetric
brackets, related to each other by higher homotopy versions of the
Jacobi identity. Hints of such structures may be traced back to higher spin gauge theories, where the closure of gauge transformations requires field-dependent gauge transformations \cite{Berends:1984rq}. They first appeared systematically
in closed bosonic string field theory, where they govern the
generalized gauge symmetries\footnote{By `generalized' we mean gauge
  symmetries which are not restricted to automorphisms of principal
  bundles. Their action may not close off-shell, or they may also have
  a ``non-linear'' (field-dependent) action on the space of fields.} and
dynamics of the theory~\cite{Zwiebach:1992ie}, both necessarily involving
infinitely-many higher brackets. In the mathematics literature, they
are known to be dual to differential graded commutative
algebras~\cite{LadaStasheff92}. In~\cite{Linfty} it was suggested that the complete data of
any classical field theory with generalized gauge symmetries fit into truncated versions of cyclic $L_\infty$-algebras
with finitely-many non-vanishing brackets, again encoding both gauge
transformations and dynamics. The existence of such an
$L_{\infty}$-algebra formulation is a consequence of the duality with the BV--BRST
formalism for perturbative field theories, which is the differential graded commutative algebra incarnation~\cite{BVChristian}. 

Previous attempts at formulating general relativity in the
$L_{\infty}$-algebra language have focused on the Einstein-Hilbert
formulation. This necessitates perturbing the metric around a fixed
background solution to the Einstein equations, which requires
infinitely-many brackets to encode the complete
dynamics~\cite{Linfty}; similar approaches are also considered
in~\cite{Reiterer:2018wcb,Nutzi:2018vkl}. In contrast, the
Einstein-Cartan-Palatini (ECP) formulation of gravity may be encoded
in finitely-many brackets without requiring perturbative
expansions. However, the standard approaches to noncommutative and
nonassociative gauge theories typically also require infinitely-many
brackets~\cite{Blumenhagen:2018kwq,Kupriyanov:2019cug}, which is
partly a
consequence of the fact that the undeformed differential does not generally
obey the usual Leibniz rule with respect to the deformed
product. Furthermore, the twisted diffeomorphism symmetry of
noncommutative gravity~\cite{NCmetricgrav} does not fit nicely into an
$L_\infty$-algebra framework involving only finitely-many
brackets. This suggests finding a suitable deformation of the notion
of an $L_\infty$-algebra to make it compatible with twisted symmetries
and to accommodate the dynamics of noncommutative and nonassociative
gauge theories. This leads to a mathematical object that we call a
\emph{braided $L_\infty$-algebra} whose definition will be sketched in
the following; further details will appear in~\cite{Inprep}.

The purpose of the present contribution is to concisely summarise the
contents of the papers~\cite{ECPLinfty,Inprep}. We
start by briefly recalling the definition of classical
Einstein-Cartan-Palatini gravity, its symmetries and its Noether
identities. We then recall the definition of cyclic
$L_{\infty}$-algebras and their applications to perturbative field
theories, and subsequently present the classical cyclic
$L_{\infty}$-algebra formulation of ECP gravity which was
developed and studied in detail in~\cite{ECPLinfty}. We proceed to describe
a braided deformation of these $L_\infty$-algebras and the
noncommutative theory of gravity that they determine, whose details will be developed in the
forthcoming paper~\cite{Inprep}. We explain how braided gauge
symmetries induce Noether identities using the language of (braided)
$L_{\infty}$-algebras, thus clarifying the physical content of braided
symmetries. We conclude with a discussion of further questions and
applications that may be explored using our formalism.

\section{General relativity in the Einstein-Cartan-Palatini formulation}
\label{sec:classicalECP}

In this section we briefly review the dynamics and symmetries of
classical general relativity in vacuum using the
Einstein-Cartan-Palatini formalism. For ease of exposition, we
restrict to the theory in dimension $d=3$, in Lorentzian
signature, and without cosmological constant. Everything we say may be
straightforwardly extended to higher dimensions $d\geq 4$ including
a cosmological constant term, and for any signature of the metric;
see~\cite{ECPLinfty} for further details.

\subsection{Field content and dynamics}

The Einstein-Cartan-Palatini formulation of gravity\footnote{This is
  also known as the Einstein-Cartan-Sciama-Kibble theory, or simply
  the first order formalism for general relativity.} on a
three-dimensional oriented manifold $M$ admitting a Lorentzian
structure is defined by the action functional
\begin{align}\label{4dAction}
S_{\textrm{\tiny ECP}}[e,\om]=\int_{M}\,\Tr (e\dwedge
  R)=\int_{M}\, \epsi_{abc} \, e^{a} \wedge R^{bc} \ .
\end{align}
The field content consists of an orientation-preserving bundle
morphism $e:TM \rightarrow \CV$ from the tangent bundle of $M$ to a
fixed oriented vector bundle $\CV$ isomorphic to $TM$ which is
endowed with a fiberwise Minkowski metric $\eta$; this defines
the coframe field $e\in\Omega^1(M,\CV)$, where $\CV$ is sometimes
called the `fake tangent bundle'. The 2-form
\begin{align}
R=\dd \om +
\tfrac{1}{2}\,[\om,\om] \ \in \ \Omega^2\big(M,P\times_{\rm
  ad}\mathfrak{so}(1,2)\big)
\end{align}
is the curvature of a connection $\om$,
called the spin connection, on
the associated principal $\sSO(1,2)$-bundle $P\to M$. The operation
$\Tr:\Omega^3(M,\midwedge^3\CV)\to\Omega^3(M)$ is induced by the natural
volume form of $\CV$.

Since every orientable three-manifold
is parallelizable, we can take $\CV$ to be the
trivial vector bundle $M\times\FR^{1,2}$. Its associated
$\sSO(1,2)$-bundle $P$ is then also trivial. In this case, the fields may
be globally identified as $1$-forms $e \in \Omega^{1}(M,\FR^{1,2})$
and $\om \in \Omega^{1}\big(M,\mathfrak{so}(1,2)\big)$. Using the isomorphism
$\mathfrak{so}(1,2)\cong \midwedge^{2} \FR^{1,2}$, one identifies the
spin connection as an element $\om \in \Omega^{1}(M,\midwedge^{2}
\FR^{1,2})$, and then the $\dwedge$-product separately takes the
exterior products of the `curved' spacetime differential forms on the
manifold $M$
and `flat' spacetime forms on the vector space $\FR^{1,2}$. Then
$\Tr:\midwedge^3\FR^{1,2}\to\FR$ is simply the Hodge duality
operator on Minkowski space $(\FR^{1,2},\eta)$.

The Euler-Lagrange derivatives which follow from varying the action
functional \eqref{4dAction} with respect to the coframe and spin connection respectively are
\begin{align}\label{4dEOM}
F_{e}:= R \qquad
          \mbox{and} \qquad
F_{\om}:= T \ ,
\end{align}
where $T$ is the torsion $2$-form which is defined as the covariant derivative of the coframe field:
\begin{align}
T:= \dd^{\om}e = \dd e + \om \wedge e \, \, \, \in \, \, \,
  \Omega^{2}(M,\FR^{1,2}) \ .
\end{align} 
The vanishing of the second equation, $ T= 0$, implies that the
torsion vanishes. When $e$ is a non-degenerate coframe field, or equivalently a bundle
isomorphism, then this equation can be solved to identify the
$\sSO(1,2)$-connection $\om$ with the Levi-Civita
connection for the metric
$g:= e^{\ast} \eta= \eta^{ab} \, e_{a}\otimes e_{b} $. The vanishing
of the first equation, $ R=0$, is then
the vacuum Einstein field equation in three dimensions, and thus the theory is
classically equivalent to general relativity. Classical solutions correspond to flat
spacetimes, which are locally isometric to Minkowski space and as such
no gravitational waves exist in three dimensions. This is of course
just the triviality of gravity in three spacetime dimensions. The only non-trivial
features lie in the potential non-trivial topology of the spacetime. 

The theory in arbitrary dimensionality $d\geq 3$ can be formulated by replacing
the coframe field $e$ with the higher powers $e^{d-2}$ in the action
functional \eqref{4dAction}, and in four dimensions the theory is
again classically
equivalent to general relativity which now possesses propagating
degrees of freedom. 
However, in contrast to
the Einstein-Hilbert formulation, the ECP theory makes sense for
degenerate coframes $e$; this extension of general relativity to a
theory with singularities is
required for its $L_\infty$-algebra formulation, where the space of fields is
required to be a vector or affine space. 

\subsection{Gauge symmetries and Noether identities}

The (infinitesimal) gauge symmetries of the action functional
\eqref{4dAction} are given
by the semi-direct product of the Lie algebras of vector fields on $M$
and of local Lorentz rotations:
\begin{align}\label{ClassicalSymmetries}
\Gamma(TM)\ltimes \Omega^{0}\big(M,\mathfrak{so}(1,2)\big) \ .
\end{align}
The diffeomorphism part is the standard symmetry of general
relativity. The local rotation part may be interpreted as the freedom
to change between orthonormal coframes for a given metric. The action
of $(\xi,\rho)\in\Gamma(TM)\ltimes \Omega^{0}(M,\mathfrak{so}(1,2))$
on the space of fields is given by
\begin{align}\label{symmetryaction}
\delta_{(\xi,\rho)}(e,\om):=(\LL_{\xi} e - \rho\cdot e\,,\, \LL_{\xi} \om +\dd^{\om}\rho)
\end{align}
where $\LL_\xi$ is the Lie derivative along the vector field $\xi$,
$\rho\cdot e$ denotes matrix multiplication,
and $\dd^{\om}\rho = \dd \rho + [\om,\rho]$ is the covariant
derivative.

By Noether's second theorem, gauge symmetries of an action functional
are in bijection with (off-shell) differential identities between its
Euler-Lagrange derivatives, called Noether identities. In the present case, these correspond to
the pair of  differential identities among the Euler-Lagrange
derivatives $(F_{e},F_{\om})$ given by
\begin{align}\label{eq:ECPNoether}
\dd x^\mu\otimes\Tr \big(
\iota_\mu \dd
e \dwedge F_{e} +
\iota_\mu \dd\om
\dwedge F_{\om}   
-\iota_\mu e
\dwedge \dd F_{e} -\iota_\mu \om\dwedge
\dd
F_{\om}\big) &= 0 \ \in \ \Omega^1\big(M,\Omega^3(M)\big) \ , \nn
  \\[4pt]
-F_e\wedge e+ \dd^\om F_\om &= 0 \ \in \ \Omega^3(M,\FR^{1,2}) \ ,
\end{align}
where $\iota_\mu$ denotes the contraction with vectors
$\partial_\mu=\frac\partial{\partial x^\mu}$ of the 
local holonomic frame dual to the basis $\{\dd x^\mu\}$ of one-forms
in a local coordinate chart on $M$, and
we identify the vector space of $1$-forms valued in $3$-forms $\Omega^1\big(M,\Omega^3(M)\big) $ with
$\Omega^1(M) \otimes\Omega^3(M)$.
The first identity is the Noether identity
corresponding to local diffeomorphism invariance
$\delta_{(\xi,0)} S_{\textrm{\tiny ECP}}[e,\om]=0$. The second identity gives the Noether
identity corresponding to the local Lorentz gauge symmetry
$\delta_{(0,\rho)} S_{\textrm{\tiny ECP}}[e,\om]=0$, which also follows from the geometrical first Bianchi identity $\dd^{\om}T=R\wedge e$ by applying the covariant
derivative to $F_{\om}$.
In three dimensions the first Bianchi identity coincides identically
with the Noether identity for local Lorentz transformations, while in higher
dimensions it only implies the Noether identity. 

\section{$L_{\infty}$-algebras of classical gravity}

In this section we review the definition of cyclic
$L_{\infty}$-algebras, and how classical field theories are formulated using them.
We then present the $L_\infty$-algebras which organise ECP gravity in
three dimensions. 

\subsection{Cyclic $L_{\infty}$-algebras }

An $L_\infty$-algebra is a $\RZ$-graded vector space $V=\bigoplus_{k\in \RZ}\, V_{k}$ equipped with graded antisymmetric multilinear maps
\begin{align}
\ell_n: \text{\Large$\otimes$}^{n}V \longrightarrow V \ , \quad  v_1\otimes \cdots\otimes v_n \longmapsto \ell_n (v_1,\dots,v_n)
\end{align}
for each $n\geq1$, of degrees $|\ell_n|=2-n$, called $n$-brackets. The graded antisymmetry translates to 
\begin{equation}\label{eq:gradedantisym}
\ell_n (\dots, v,v',\dots) = -(-1)^{|v|\,|v'|}\, \ell_n (\dots, v',v,\dots) \ ,
\end{equation}
where we denote the degree of a homogeneous element $v\in V$ by
$|v|$. Degreewise, the $n$-brackets are thus maps
$\ell_n:V_{k_1}\otimes\cdots\otimes V_{k_n}\to V_{k_1+\cdots+k_n+2-n}$.

The $n$-brackets $\ell_n$ are required to fulfill infinitely-many
identities, called homotopy relations, for each $n\geq1$:
\begin{align} \label{eq:calJndef}
\sum^n_{i=1}\, (-1)^{i\,(n-i)} \
\sum_{\sigma\in{\rm Sh}_{i,n-i}} \, \chi(\sigma;v_1,\dots,v_n) 
\,
\ell_{n+1-i}\big(
\ell_i(v_{\sigma(1)},\dots,
v_{\sigma(i)}),
v_{\sigma(i+1)},\dots
,v_{\sigma(n)}\big) = 0
\ , 
\end{align}
where, for each $i=1,\dots,n$, the second sum runs over $(i,n-i)$-shuffled permutations
$\sigma\in S_n$ of degree $n$ which are restricted as 
$\sigma(1)<\dots<\sigma(i)$ and $\sigma(i+1)<\dots < \sigma(n)$.
The Koszul sign $\chi(\sigma;v_1,\dots,v_n)=\pm\, 1$ is determined from the grading by
\begin{align}
v_{\sigma(1)}\wedge\cdots\wedge v_{\sigma(n)} = \chi(\sigma;v_1,\dots,v_n) \ v_1\wedge\cdots\wedge v_n \ .
\end{align}

Let us examine the first three identities explicitly. The first
relation for $n=1$ is
\begin{align}
\ell_1\big(\ell_1(v)\big) = 0 \ ,
\end{align}
which states that the map $\ell_1:V\to V$ is a
differential making $(V,\ell_1)$ into a cochain complex. The second
relation with $n=2$ reads
\begin{align}
\ell_1\big(\ell_2(v_1,v_2)\big) = \ell_2\big(\ell_1(v_1),v_2\big) +
  (-1)^{|v_1|}\, \ell_2\big(v_1, \ell_1(v_2)\big) \ , 
\end{align}
which states that $\ell_1$ is a graded
derivation of the $2$-bracket $\ell_2$. The third relation for $n=3$
is
\begin{align}
&\ell_2\big(\ell_2(v_1,v_2),v_3\big) + (-1)^{(|v_1|+|v_2|)\,|v_3|}\,
  \ell_2\big(\ell_2(v_3,v_1),v_2\big) +  (-1)^{(|v_2|+|v_3|)\,|v_1|}\,
  \ell_2\big(\ell_2(v_2,v_3),v_1\big) \nn\\[4pt]
& \qquad = - \ell_1\big(\ell_3(v_1,v_2,v_3)\big) \label{I3}\\
& \qquad \qquad - \ell_3\big(\ell_1(v_1),v_2,v_3\big) - (-1)^{|v_1|}\, \ell_3\big(v_1, \ell_1(v_2), v_3\big) - (-1)^{|v_1|+|v_2|}\, \ell_3\big(v_1,v_2, \ell_1(v_3)\big) \ , \nn
\end{align}
which states that the graded Jacobi identity for the
$2$-bracket $\ell_2$ holds generally only up to
coherent homotopy. Thus $L_\infty$-algebras are (strong) homotopy
deformations of differential graded Lie algebras which are the special
cases where the ternary and all higher brackets vanish: $\ell_n=0$ for
all~$n\geq3$. In general, the homotopy relations for $n\geq3$ are
generalized Jacobi identities.

A cyclic $L_{\infty}$-algebra is an $L_{\infty}$-algebra $V$ equipped
with a non-degenerate graded symmetric bilinear pairing 
\begin{align}
\langle -,-\rangle : V\times V \longrightarrow \FR
\end{align}
which satisfies the cyclicity property 
\begin{align}
\langle v_0,\ell_n(v_1,v_2,\dots,v_n)\rangle =
  -(-1)^{n+(|v_0|+|v_n|)\,n+|v_n| \,\sum_{i=0}^{n-1}\,|v_i|} \ \langle
  v_n,\ell_n(v_0,v_1,\dots,v_{n-1})\rangle \ .
\end{align} 
This is the natural generalization of the notion of an invariant inner
product on a Lie algebra. Thus cyclic $L_\infty$-algebras generalize
quadratic Lie algebras.

\subsection{$L_{\infty}$-algebra formulation of classical field
  theories}

Given a classical field theory defined by a (polynomial) action
functional $S$ on a vector (or affine) space of dynamical fields with an
irreducible set of generalized gauge symmetries, there is a $4$-term
$L_{\infty}$-algebra with underlying graded vector space given by
\begin{align}
V=V_{0}\oplus V_{1}\oplus V_{2} \oplus V_3 \ ,
\end{align}
where the different subspaces in degrees $0$, $1$, $2$ and $3$ contain the gauge parameters,
the dynamical fields, the Euler-Lagrange derivatives and the Noether identities,
respectively. The linear parts of the gauge transformations, field
equations and Noether identities are encoded by a differential
$\ell_{1}:V\rightarrow V$, yielding a cochain complex $(V,\ell_1)$. 
This complex is further equipped with suitable
higher brackets $\ell_n$ corresponding to the non-linear parts of the theory subject to the homotopy relations in order the recover
the full symmetries and dynamics of the generalized gauge theory.

Given $\lambda \in V_{0}$ and $A\in V_{1}$, the gauge
variations are encoded as the maps $A\mapsto A+\delta_\lambda A$ where
\begin{align} \label{gaugetransfA}
\delta_{\lambda}A=\sum_{n =0}^\infty \, \frac{1}{n!}\, (-1)^{\frac12\,{n\,(n-1)}}\, \ell_{n+1}(\lambda,A,\dots,A) \ ,
\end{align}
and the brackets involve $n$ insertions of the dynamical field
$A$. 
The Euler-Lagrange derivatives are encoded as
\begin{align}\label{EOM} 
F_A=\sum_{n =1}^\infty \, \frac{1}{n!}\, (-1)^{\frac12\,{n\,(n-1)}}\, \ell_{n}(A,\dots,A) \ ,
\end{align}
with the covariant gauge variations
\begin{align} \label{gaugetransfF}
\delta_{\lambda} F_A=\sum_{n =0}^\infty \, \frac{1}{n!}\, (-1)^{\frac12\,{n\,(n-1)}}\, \ell_{n+2}(\lambda,F_A,A,\dots,A) \ .
\end{align}
If the distribution in $TV_1$ spanned by the gauge parameters is
involutive off-shell, that is, when $F_A\neq0$, then the closure relation for the gauge algebra has the form
\begin{align}\label{eq:closure}
[\delta_{\lambda_1},\delta_{\lambda_2}]A
= \delta_{[\![\lambda_1,\lambda_2]\!]_A}A \ ,
\end{align}
where 
\begin{align}
[\![\lambda_1,\lambda_2]\!]_A = -\sum_{n=0}^\infty \, \frac1{n!}\,
(-1)^{\frac12\,{n\,(n-1)}}\, \ell_{n+2}(\lambda_1,\lambda_2,A,\dots,A)
\ .
\end{align}
The homotopy relations guarantee that the Jacobi
identity is generally satisfied for any triple of maps $\delta_{\lambda_1}$,
$\delta_{\lambda_2}$ and~$\delta_{\lambda_3}$.
Finally, the Noether identities are encoded by
\begin{align}\label{eq:Noether}
\sum_{n=0}^\infty \, \frac1{n!} \, (-1)^{\frac12\, n\, (n-1)} \, \ell_{n+1}(F_A,A,\dots,A) =0 \ ,
\end{align}
which vanishes identically (off-shell) as a consequence of the
homotopy relations \eqref{eq:calJndef} with all entries set equal to $A$. 

To encode the action functional of the gauge field theory, we assume
that $V$ is equipped with a cyclic pairing of degree~$-3$, which makes $V$ into a cyclic $L_\infty$-algebra. 
In this case it is easy to see that the field equations $F_A = 0$ follow from varying the action functional defined as 
\begin{align} \label{action}
S[A] := \sum_{n=1}^\infty \, \frac{1}{(n+1)!}\, (-1)^{\frac12\,{n\,(n-1)}}\, \langle A, \ell_{n}(A,\dots,A)\rangle \ ,
\end{align}
since then cyclicity implies $\delta S[A]=\langle F_A,\delta
A\rangle$. Gauge invariance of the action functional, $0=\delta_\lambda
S[A]=\langle F_A,\delta_\lambda A\rangle$, together with cyclicity on
$V_0$ then imply the Noether identities \eqref{eq:Noether}.

\subsection{$L_{\infty}$-algebra picture of Einstein-Cartan-Palatini
  gravity}
\label{sec:LinftyECP}

We shall now present the $L_\infty$-algebra for ECP gravity in
dimension $d=3$. The story may be extended to any $d\geq 3$, any
signature, and incorporating a cosmological constant. For further
details, along with explicit proofs of the homotopy relations and
cyclicity conditions, we point the interested reader to
\cite{ECPLinfty}.

The $L_{\infty}$-algebra which organises the gravity theory in three
dimensions from Section~\ref{sec:classicalECP} reduces to a
differential graded Lie algebra, that is, the brackets $\ell_n$ vanish
for all $n \geq 3$. The underlying cochain complex is given by
\begin{align}\label{eq:3dV}
V_0\xrightarrow{ \ \ \ell_1 \ \ }V_1\xrightarrow{ \ \ \ell_1 \ \ 
}V_2\xrightarrow{ \ \ \ell_1 \ \ }V_3 
\end{align}
where 
\begin{align} 
V_{0}&=\Gamma(TM)\times \Omega^{0}\big(M,\mathfrak{so}(1,2)\big) \ , \nn
\\[4pt]
V_{1}&= \Omega^{1}(M,\FR^{1,2}) \times
\Omega^{1}\big(M,\mathfrak{so}(1,2)\big) \ , \label{eq:3dV012} \\[4pt]
V_{2}&=\Omega^{2}\big(M,\midwedge^{2}(\FR^{1,2})\big) \times
\Omega^{2}(M,\FR^{1,2}) \ , \nn \\[4pt]
V_3&= \Omega^1\big(M,\Omega^3(M)\big) \times \Omega^3(M,\FR^{1,2})
\ . \nn
\end{align} 
We denote elements of these vector spaces by $(\xi,\rho)\in V_0$,
$(e,\om)\in V_1$, $(E,{\mit\Omega})\in V_2$ and $(\CX,\CP)\in V_3$. The
differential is given by
\begin{align} 
\ell_{1}(\xi,\rho)=(0,\dd\rho) \ , \quad
\ell_{1}(e,\omega)=(\dd \om,\dd e) \qquad \mbox{and} \qquad
\ell_{1}(E,{\mit\Omega})=(0,\dd {\mit\Omega}) \ .
\end{align}
The non-trivial $2$-brackets are
\begin{flalign}
\ell_{2}\big((\xi_{1},\rho_{1})\,,\,(\xi_{2},\rho_{2})\big)&=\big([\xi_{1},\xi_{2}]\,,\,-[\rho_{1},\rho_{2}]+\xi_1(\rho_{2})
- \xi_{2}(\rho_{1})\big) \ , \nn
\\[4pt]
\ell_{2}\big((\xi,\rho)\,,\,(e,\omega)\big)&=\big(-\rho \cdot e
+\LL_{\xi}e \,,\, -[\rho,\omega]+\LL_{\xi}\om\big) \
, \label{eq:3dbrackets} \\[4pt]
\ell_{2}\big((\xi,\rho) \,,\, (E,{\mit\Omega})\big)&=\big(-
[\rho, E]+\LL_{\xi}E \,,\, -\rho \cdot
{\mit\Omega}+\LL_{\xi}{\mit\Omega} \big) \ , \nn \\[4pt]
\ell_{2}\big((\xi,\rho)\,,\,({\CX},{\CP})\big)&= \big
(\dd x^\mu\otimes\Tr(\iota_\mu \dd \rho \dwedge
{\CP}) +\LL_{\xi}{\CX}\,,\,-\rho
\cdot {\CP} +\LL_{\xi} {\CP}\big) \ , \nn \\[4pt] 
\ell_{2}\big((e_{1},\omega_{1})\,,\,(e_{2},\omega_{2})\big)&=-\big([\omega_{1}, \omega_{2}]
\,,\, \omega_{1}\wedge
e_{2} +\omega_{2} \wedge e_{1} \big) \ , \nn \\[4pt]
\ell_{2}\big((e,\om)\,,\,(E,{\mit\Omega})\big)&=\big(\dd x^\mu\otimes\Tr ( \iota_\mu \dd e \dwedge E + \iota_\mu \dd\om \dwedge {\mit\Omega} \nn   -\iota_\mu e \dwedge \dd E - \iota_\mu \om\dwedge \dd {\mit\Omega}) \,, \nn \\ 
&\hspace{8cm}
E\wedge e - \omega \wedge {\mit\Omega} \big)  \ . \nn
\end{flalign}

As designed by
\eqref{gaugetransfA}--\eqref{eq:Noether}, these brackets encode the
symmetries and dynamics of three-dimensional gravity. In summary:
\begin{myitemize}
\item $\underline{\text{Gauge transformations of fields:}}$
\begin{align}
\delta_{(\xi,\rho)}(e,\om) = (-\rho\cdot e+\LL_\xi
  e\,,\,\dd\rho-[\rho,\om]+\LL_\xi\om) =
  \ell_1(\xi,\rho)+\ell_2\big((\xi,\rho)\,,\,(e,\om)\big) \ .
\end{align}
\item $\underline{\text{Euler-Lagrange derivatives:}}$
\begin{align}
F_{(e,\omega)} = (R,T) 
=(\dd \om,\dd e)+ \tfrac{1}{2} \, ([\omega,\omega],2\,\omega\wedge e)
=\ell_{1}(e,\omega)-\tfrac{1}{2}\,
\ell_{2}\big((e,\omega)\,,\,(e,\omega)\big) \ .
\end{align}
\item $\underline{\text{Covariance of field equations:}}$
\begin{align}
\delta_{(\xi,\rho)}F_{(e,\om)} = (-[\rho,R]+\LL_\xi R,-\rho\cdot
  T+\LL_\xi T) = \ell_2\big((\xi,\rho)\,,\,F_{(e,\om)}\big) \ .
\end{align}
\item $\underline{\text{Noether identities:}}$
\begin{align}
&\big(\dd x^\mu\otimes\Tr(\iota_\mu e\dwedge\dd
  R-\iota_\mu\dd e\dwedge R +
  \iota_\mu\om\dwedge\dd T -
  \iota_\mu\dd\om\dwedge T)\,,\,\dd^\om T-R\wedge e\big)
  \nn \\[4pt]
& \hspace{3cm}
  = \ell_1(F_{(e,\om)})-\ell_2\big((e,\om)\,,\,F_{(e,\om)}\big) \nn
  \\[4pt]
& \hspace{5cm} = (0,0) \ .
\end{align}
\end{myitemize}

To write the action functional \eqref{4dAction} in this language, we
define a cyclic pairing of degree $-3$ by
\begin{align} \label{eq:ECPpairing}
\langle (e,\om) \,,\, (E,{\mit\Omega}) \rangle:= 
\int_{M}\, \Tr \big(e\dwedge E+ {\mit\Omega} \dwedge \om \big)
\end{align}
on $V_{1} \times V_{2}$. It is extended by
\begin{align} \label{eq:ECPpairing2}
\langle(\xi,\rho)\,,\,({\CX},{\CP})\rangle := \int_M\,
\iota_\xi{\CX} + \int_M\, \Tr\big(\rho\dwedge{\CP}\big)
\end{align}
on $V_0\times V_3$, where $\iota_\xi$ denotes contraction with the
vector field $\xi$. Then the ECP action functional can be written as in \eqref{action} using the cyclic pairing~\eqref{eq:ECPpairing}:
\begin{align}
S_{\textrm{\tiny ECP}}[e,\omega] &= \int_M\, \Tr
\Big(e\dwedge\big(\dd\omega+\tfrac12\, [\omega,\omega]\big)\Big) \nn \\[4pt]
&= \tfrac12\, \big\langle (e,\omega)\,,\,\ell_1(e,\omega) \big\rangle -
\tfrac1{6}\, \big\langle (e,\omega)\,,\,\ell_2\big(
(e,\omega)\,,\,(e,\omega)\big) \big\rangle \ .
\end{align}

One application of this formulation gives a new (and deeper)
perspective on the well-known
equivalence between three-dimensional gravity (with vanishing
cosmological constant) and Chern-Simons theory
with gauge algebra
$\mathfrak{iso}(1,2)=\FR^3\rtimes\mathfrak{so}(1,2)$, the Lie algebra
of the Poincar\'e group in three dimensions~\cite{Witten:1988hc}. The
ECP theory in three dimensions possesses an extra ``translation''
symmetry, whose Noether identity is the second Bianchi identity
$\dd^\om R=0$, and its $L_\infty$-algebra may be extended to include this
new gauge symmetry. In the case of non-degenerate coframes (metrics)
one can show that the gauge orbits under (infinitesimal)
diffeomorphisms and ``translations'' coincide on-shell. However, if one
allows for degenerate metrics then the ``translation'' gauge orbits
strictly include the diffeomorphism orbits on-shell, thus rendering
the diffeomorphisms as redundant symmetries. This may be used to show that
the differential graded Lie algebra of ECP gravity including
degenerate coframes is quasi-isomorphic (as an $L_\infty$-algebra) to the
differential graded Lie algebra of Chern-Simons theory, thus extending
the equivalence in an off-shell sense. See~\cite{ECPLinfty} for
further details and applications.

In higher dimensions $d\geq4$, the ECP $L_\infty$-algebra contains
higher brackets and is no longer simply a differential graded Lie
algebra, owing to the higher degree polynomial nature of the action
functional (in the coframes) and the field
equations. In~\cite{ECPLinfty} is shown that, for any spacetime dimension, these $L_{\infty}$-algebras
describe the complete BV--BRST formulation of ECP gravity; in
particular, in four dimensions
our $L_\infty$-algebras are dual to the BV--BRST formulation of~\cite{ECBV}.

\section{Braided $L_\infty$-algebras of noncommutative gravity}

In this section we explain how suitable Drinfel'd twists give rise to
a braided noncommutative deformation of the ECP gravity theory and its
$L_{\infty}$-algebras. Again for brevity we state everything in three
spacetime dimensions, which captures the main novelties of our braided
gauge symmetry approach. The cases of higher spacetime dimensions,
together with proofs of the claims which follow and further
explanations, will appear in the forthcoming paper~\cite{Inprep}.

\subsection{Drinfel'd twist deformations of manifolds}

We recall how to twist a manifold to a noncommutative space
using a Drinfel'd twist. Let $M$ be a manifold and consider its
Lie algebra of diffeomorphisms $\Gamma(TM)$. The universal enveloping algebra
$U\Gamma(TM)$ is the tensor algebra (over $\FR$) of $\Gamma(TM)$,
regarded as the free unital algebra generated by $\Gamma(TM)$,
modulo the two-sided ideal generated by $\xi_1\, \xi_2 -\xi_2\, \xi_1-
[\xi_1,\xi_2]$ for all $\xi_1,\xi_2\in\Gamma(TM)$. It is
naturally a cocommutative Hopf algebra with the coproduct $\Delta:U\Gamma(TM)\to U\Gamma(TM)\otimes
U\Gamma(TM)$ given by 
\begin{align}\label{eq:ClassicalHopf}
\Delta(\xi)&= \xi\otimes 1 + 1\otimes \xi\qquad \mbox{and} \qquad \Delta(1)=1\otimes 1 
\end{align}
for all $\xi\in \Gamma(TM)$, and extended as an algebra homomorphism to the whole of $U\Gamma(TM)$.

There is a symmetric monoidal category whose objects are
$U\Gamma(TM)$-modules and whose morphisms are equivariant maps (see e.g.~\cite{Barnes:2014ksa}).
A $U\Gamma(TM)$-module algebra is an algebra object in this category,
that is, an algebra $\CA$ with a
$U\Gamma(TM)$-action $\triangleright: U\Gamma(TM)\otimes \CA
\rightarrow \CA$ which is compatible with the algebra multiplication
via the coproduct $\Delta$. In the following the main example of
interest will be the exterior algebra of
differential forms $\CA =\big(\Omega^{\bullet}(M),\wedge\big)$, which
carries a $\Gamma(TM)$-action via the Lie derivative, extending to
a $U\Gamma(TM)$-action via successive applications of the Lie
derivative. This is a $U\Gamma(TM)$-module algebra since
\begin{align}
\xi \triangleright (\alpha\wedge \beta) &= \LL_{\xi}(\alpha\wedge \beta)=\LL_{\xi} \alpha \wedge \beta+ \alpha \wedge \LL_{\xi} \beta =\wedge \circ \Delta(\xi) \triangleright(\alpha\otimes \beta)
\end{align}
for all $\xi\in\Gamma(TM)$ and $\alpha,\beta\in
\Omega^{\bullet}(M)$. In the following we will usually drop the symbol
$\triangleright$ to simplify the notation.

A Drinfel'd twist is a (normalized) 2-cocycle of the Hopf algebra
$U\Gamma(TM)$, which is specified by an invertible element $\CF \in U\Gamma(TM) \otimes U\Gamma(TM)$. In most applications
the twists are actually elements of formal power series expansions $\CF=:\sff^{k}\otimes
\sff_{k}$ in
a deformation parameter, with all spaces concerned also similarly
extended, but we do not indicate this in the notation for
simplicity. For example, on $M=\FR^d$ the standard example is the
abelian Hermitean Moyal-Weyl twist
\begin{align}
\CF =
  \exp\big(-\tfrac{\mathrm{i}}2\,\theta^{\mu\nu}\,\partial_\mu\otimes\partial_\nu\big)
  \ ,
\end{align}
where $(\theta^{\mu\nu})$ is a $d{\times}d$ antisymmetric real
matrix. However, our considerations in the following apply to a more
general class of Drinfel'd twists that we shall specify momentarily.

A Drinfel'd twist defines a new Hopf algebra structure on the
universal enveloping algebra, which we denote by
$U_{\CF}\Gamma(TM)$. It has a twisted coproduct 
\begin{align}\label{eq:TwistedHopf}
\Delta_{\CF}(X):= \CF \, \Delta(X) \, \CF^{-1}
\end{align}
for all $X\in U\Gamma(TM)$, where $\CF^{-1}=: \bar{\sff}^{k}\otimes
\bar{\sff}_{k}$. The deformation map defines a functorially equivalent symmetric monoidal category of
$U_\CF\Gamma(TM)$-modules. If $\CA$ is a $U\Gamma(TM)$-module algebra, then we
can deform the product on $\CA$ by precomposing it with
$\CF^{-1}$. The cocycle condition on $\CF$ guarantees that this
defines an associative star-product, and it generally defines a
noncommutative $U_\CF\Gamma(TM)$-module algebra $\CA_\star$, that is, an algebra object
in the category of $U_\CF\Gamma(TM)$-modules; if $\CA$ is commutative,
then $\CA_\star$ is braided commutative. Let us spell this out
explicitly on our main example of interest
$\CA=\big(\Omega^\bullet(M),\wedge)$. For $\alpha,\beta\in
\Omega^{\bullet}(M)$ we set
\begin{align}
\alpha\wedge_\star\beta:=\wedge\circ\CF^{-1}(\alpha\otimes\beta) =
\bar\sff^k\alpha \wedge \bar\sff_k\beta \ ,
\end{align}
with graded commutativity controlled up to a braiding given by the
triangular $\RR$-matrix 
\begin{align}
\RR=\CF_{21}\, \CF^{-1}=:\sfR^k\otimes\sfR_k
\end{align}
where $\CF_{21}=\sff_k\otimes\sff^k$ is the twist with its legs
flipped. Explicitly
\begin{align}
\alpha\wedge_{\star} \beta = (-1)^{|\alpha| \,|\beta|} \ \bar{\sfR}^{k} \beta
  \wedge_{\star} \bar{\sfR}_{k} \alpha \ ,
\end{align}
where $\RR^{-1}=:\bar{\sfR}^k\otimes\bar{\sfR}_k$.
The original action $\triangleright: U\Gamma(TM)\otimes \CA \rightarrow \CA$ is now compatible with the star-product via the twisted coproduct $\Delta_{\CF}$:
\begin{align}
X \triangleright(\alpha\wedge_\star \beta )= \wedge_{\star}\circ
  \Delta_{\CF}(X)(\alpha\otimes \beta) \ ,
\end{align} 
and the twisted exterior algebra $\CA_\star=\big(\Omega^{\bullet}(M),\wedge_{\star}\big)$ is
now a $U_{\CF}\Gamma(TM)$-module algebra. For further details on Drinfel'd
twists see e.g. \cite{MajidBook}, and~\cite{Szabo:2006wx} for a review of their
applications to twisted symmetries.

We will see below that $L_\infty$-algebras can be twisted in a similar manner, and thus twist the
kinematics, symmetries and dynamics of classical field theories
simultaneously in a consistent way. More generally, if $\CF$ is a cochain twist (dropping the cocycle condition), then $U_\CF\Gamma(TM)$ is
a quasi-Hopf algebra and $\CA_\star$ is a nonassociative algebra. This
is the path towards a nonassociative theory of gravity taken
in~\cite{Mylonas:2013jha,Barnes:2014ksa,Aschieri:2015roa,Blumenhagen:2016vpb,Aschieri:2017sug,Barnes:2015uxa,Barnes:2016cjm},
where further details of the cochain twist deformation may be found. However,
while this is in principle a straightforward generalization, in the
following we shall stick to the strictly associative noncommutative
case for simplicity.

\subsection{Braided Einstein-Cartan-Palatini gravity}

\paragraph{Braided gauge symmetries and kinematics.}
One way to introduce a noncommutative theory of gravity is to start
with the classical symmetries $\Gamma(TM)\ltimes
\Omega^{0}\big(M,\mathfrak{so}(1,2)\big)$ and deform them
consistently. That is, for any Drinfel'd twist $\CF$, we define the
\emph{braided Lie algebra} structure given by the brackets
\begin{align}
[\rho_{1},\rho_{2}]_{\star}:=[-,-]\circ
  \CF^{-1}(\rho_{1}\otimes\rho_{2}) 
\end{align}
for $\rho_{1},\rho_{2} \in \Omega^{0}\big(M,\mathfrak{so}(1,2)
\big)$. The new bracket is now braided antisymmetric: 
\begin{align}
[\rho_{1},\rho_{2}]_{\star}=-[\bar{\sfR}^{k}\rho_{2},\bar{\sfR}_{k}\rho_{1}]_\star
  \ ,
\end{align}
and satisfies the braided Jacobi identity:
\begin{align}
[\rho_{1},[\rho_{2},\rho_{3}]_{\star}]_{\star}=[[\rho_{1},\rho_{2}]_{\star},\rho_{3}]_{\star}+[\bar{\sfR}^{k}\rho_{2},[\bar{\sfR}_{k}\rho_{1},\rho_{3}]_{\star}]_{\star}
  \ ,
\end{align}
for all $\rho_{1},\rho_{2},\rho_{3} \in
\Omega^{0}\big(M,\mathfrak{so}(1,2)\big)$. Similarly, one defines the braided
Lie algebra of vector fields on $M$.

This differs from the usual approach to noncommutative gauge
theory in the following way. In the standard approach one considers some matrix Lie algebra
$\frg$, and deforms the associative matrix product of functions in the
algebra $\Omega^{0}(M,\frg)$. Then the star-bracket is defined as
$[f\,\overset{\star}{,}\,g]:= f\star g - g\star f$, which is subsequently
required to close on $\Omega^0(M,\frg)$ as an
ordinary Lie algebra. This approach often requires an extension of the
set of gauge symmetries, and it was first applied to gravity
in~\cite{Chamseddine:2000si,Cardella:2002pb}. This is the approach
taken in \cite{AschCast} for ECP gravity, which necessarily introduces
new degrees of freedom into the theory.

Instead, in our approach the braided bracket closes by definition on
the same vector space, and hence avoids the introduction of new
degrees of freedom. In terms of matrix multiplication, one can see
that $[\rho_{1},\rho_{2}]_{\star}= \rho_{1}\star \rho_{2} -
\bar{\sfR}^k \rho_{2} \star \bar{\sfR}_{k} \rho_{1}$. Furthermore, the
only coherent way of deforming $\Gamma(TM)$ simultaneously is as a
braided gauge symmetry. Treating the two symmetries in a democratic
way has positive outcomes. For example, the two braided Lie algebras may be combined in a
single (braided) semi-direct product $\Gamma(TM)\ltimes_{\star}
\Omega^{0}\big(M,\mathfrak{so}(1,2)\big)$, preserving the classical
property of the gauge symmetries.

Furthermore, representations of the classical Lie algebras induce
braided representations. In particular, we may define a notion of a braided
coframe field $e\in \Omega^{1}(M,\FR^{1,2}) $ and of a braided spin connection $\om\in
\Omega^{1}\big(M,\mathfrak{so}(1,2)\big)$ which transform as
\begin{align}\label{eq:braidedtransf}
\delta_{\rho}^{\star}e=-\rho \star e \qquad \mbox{and} \qquad
  \delta_{\rho}^{\star}\om= \dd \rho - [\rho,\om]_{\star}= \dd \rho
  -\rho\star \om+\bar{\sfR}^{k}\om\star \bar{\sfR}_{k}\rho \ ,
\end{align}
where the products are defined by matrix
multiplication with the star-product. These braided gauge
transformations satisfy the braided Leibniz rule, for example
\begin{align}\label{eq:braidedleibniz}
\delta_{\rho}^{\star}(e\otimes \om)=\delta_{\rho}^{\star}e\otimes \om
  +\bar{\sfR}^{k} e\otimes \delta^\star_{\bar{\sfR}_{k}\rho}\om \ .
\end{align}
However, the exterior derivative $\dd$ itself acts via the ordinary,
undeformed Leibniz rule.
The braided torsion and braided curvature are naturally defined as
\begin{align}\label{eq:braidedTorsCurv}
T:=\dd e +\om \wedge_{\star}e \qquad \mbox{and} \qquad R:= \dd \om
  +\tfrac{1}{2}\,[\om,\om]_{\star} \ ,
\end{align}
which as expected transform covariantly under (\ref{eq:braidedtransf})
and (\ref{eq:braidedleibniz}). 

The action of the braided
diffeomorphisms is defined accordingly by twisting the classical
action of vector fields. For example
\begin{align}\label{eq:BraidedLie}
\delta_{\xi}^{\star}e=\LL_{\xi}^{\star}e:= \LL_{\bar{\sff}^k\xi}
  \,(\,\bar{\sff}_{k} e) 
\end{align}
 for $\xi \in \Gamma(TM)$.

\paragraph{Dynamics.}
The action functional for noncommutative ECP gravity in three
dimensions is given by~\cite{Barnes:2016cjm,Inprep}
\begin{align}\label{eq:NC3daction}
S^{\star}_{\textrm{\tiny ECP}}[e,\om]=\int_M\, \Tr  (e\dwedge_{\star}
  R)=\int_M\, \epsi_{abc} \, e^{a}\wedge_{\star} R^{bc} \ .
\end{align}
This is the unique deformation of the classical action functional
\eqref{4dAction}, given a twist $\CF$ for which the deformed exterior
product $\wedge_\star$ is (graded) cyclic commutative under the
integral; for example, this is true for abelian twists, whereby
$\CF_{21}=\CF^{-1}$. Such a graded cyclicity is
also necessary for the variational principle to make sense. Together with the assumption that $\CF$ is
Hermitean, that is, it defines a Hermitean star-product $\wedge_\star$, then ensures
that the action functional \eqref{eq:NC3daction} is real-valued. 

The action
functional \eqref{eq:NC3daction} is invariant under the action of
braided gauge transformations of 
\begin{align}
\Gamma(TM)\ltimes_\star \Omega^{0}\big(M,\mathfrak{so}(1,2)\big) \ .
\end{align}
The resulting field equations are captured by the vanishing of the
Euler-Lagrange derivatives
\begin{flalign}\label{NC3dEOM}
F_{e}^{\star}= R
\qquad \mbox{and} \qquad F_{\om}^{\star}= T-\tfrac{1}{2}\,\om\wedge_{\star}e
+\tfrac{1}{2}\,\bar{\sfR}^{k} \om \wedge_{\star} \bar{\sfR}_{k} e \ .
\end{flalign}
Thus the classical solutions may be interpreted as flat noncommutative spacetimes, however with torsion induced by the non-trivial braiding. 

One distinctive feature of braided gauge symmetries is they do not
produce new classical solutions: the field equations \eqref{NC3dEOM}
are indeed braided covariant under (\ref{eq:braidedtransf}), yet 
\begin{align}
R[\om]+\delta_{\rho}^{\star}R[\om] \neq R[\om+\delta_{\rho}^{\star}
  \om] \ ,
\end{align}
which is a consequence of the braided Leibniz rule
(\ref{eq:braidedleibniz}). Another incarnation of this is seen in the
braided gauge variations of the action functional \eqref{eq:NC3daction}:
\begin{align}
\delta_{\rho}^{\star} S_{\textrm{\tiny ECP}}^{\star}[e,\om] \neq
  \int_{M}\, \big(\delta_{\rho}^{\star}e \dwedge_{\star} F^\star_{e} +
  \delta_{\rho}^{\star}\om \dwedge_{\star} F^\star_{\om}\big) \ ,
\end{align}
in contrast to field theories based on ordinary gauge symmetries. Hence
the usual approach to Noether's second theorem also fails. However, a
braided version of Noether's second theorem may be proved which again
results in a set of a differential identities between the
Euler-Lagrange derivatives $(F_e^\star,F_\om^\star)$, that hold
off-shell; as in the commutative case, these may be derived using the
braided versions of the Bianchi identities~\cite{Inprep}. This shows that the degrees of
freedom are not independent, and justifies the interpretation of local
braided symmetries as (generalized) `gauge symmetries'. These identities are most
elegantly phrased in terms of the corresponding braided
$L_{\infty}$-algebra that we introduce below.

\subsection{Braided $L_{\infty}$-algebras}

Our definition of braided $L_{\infty}$-algebras is obtained from
classical $L_{\infty}$-algebras. Let $(V,\{\ell_{n}\})$ be an
$L_{\infty}$-algebra object in the category of $U\Gamma(TM)$-modules. This
means that the graded subspaces $V_k$ of $V$ are objects and the $n$-brackets $\ell_n$ define morphisms in this
category, which are equivariant maps; concretely, this boils down to
the action of $\Gamma(TM)$ on $V$ commuting with each of the $n$-brackets,
via the trivial coproduct $\Delta$. A Drinfel'd twist $\CF$ then induces an $L_{\infty}$-algebra object $(V,\{\ell_{n}^{\star}\})$ in the category of $U_{\CF}\Gamma(TM)$-modules, where $\ell_{1}^{\star}:=\ell_{1}$ and
\begin{align}\label{eq:braided brackets}
\ell_{n}^{\star}(v_{1}\otimes\cdots\otimes v_{n}):= \ell_{n}(v_{1}\otimes_{\star}\cdots \otimes_{\star} v_{n})
\end{align}
for $n\geq2$, with $v\otimes_\star v':= \CF^{-1}(v\otimes v')$ for
$v,v'\in V$. We call $(V,\{\ell_n^\star\})$ a \emph{braided $L_{\infty}$-algebra}, since its brackets are braided (graded) antisymmetric:
\begin{align}\label{eq:braidedantisym}
\ell_{n}^\star(\dots,v,v',\dots)=- (-1)^{|v|\, |v'|} \,
  \ell_{n}^\star(\dots, \bar{\sfR}^{k} v', \bar{\sfR}_{k} v,\dots) \ ,
\end{align}
and satisfy the braided version of the homotopy relations
\eqref{eq:calJndef}. The first two relations for $n=1,2$ read as
classically, while for $n=3$ the homotopy Jacobi identity \eqref{I3}
is modified to
\begin{align}
&\ell^\star_2\big(\ell^\star_2(v_1,v_2),v_3\big) - (-1)^{|v_2|\,|v_3|}\,
  \ell^\star_2\big(\ell^\star_2(v_1,\bar{\sfR}^kv_3),\bar{\sfR}_kv_2\big) +  (-1)^{(|v_2|+|v_3|)\,|v_1|}\,
  \ell^\star_2\big(\ell^\star_2(\bar{\sfR}^kv_2,\bar{\sfR}^lv_3),\bar{\sfR}_l\bar{\sfR}_kv_1\big) \nn\\[4pt]
& \qquad = - \ell^\star_1\big(\ell^\star_3(v_1,v_2,v_3)\big) \label{I3NC}\\
& \qquad \qquad - \ell^\star_3\big(\ell^\star_1(v_1),v_2,v_3\big) - (-1)^{|v_1|}\, \ell^\star_3\big(v_1, \ell^\star_1(v_2), v_3\big) - (-1)^{|v_1|+|v_2|}\, \ell^\star_3\big(v_1,v_2, \ell^\star_1(v_3)\big) \ , \nn
\end{align}
and similarly for all generalized Jacobi identities with $n \geq
3$. The essential feature is that the permutation action in the
identities \eqref{eq:calJndef} is enhanced by the application of the $\RR$-matrix.

Similarly, cyclic pairings $\langle-,-\rangle:V\times V\to\FR$ in the category of $U\Gamma(TM)$-modules
induce braided cyclic pairings via 
\begin{align}
\langle -, - \rangle_{\star} := \langle -, - \rangle \circ \CF^{-1} \
  .
\end{align}
However, for our choices of twists and pairings in field theory these
become \emph{strictly cyclic}.

Let us now write down the braided $L_{\infty}$-algebra of
noncommutative gravity. For this, we first observe
that the classical cyclic $L_\infty$-algebra for three-dimensional gravity from Section~\ref{sec:LinftyECP} is indeed an
object in the category of $U\Gamma(TM)$-modules, and hence we obtain a
cyclic braided $L_{\infty}$-algebra via twisting. The underlying
cochain complex \eqref{eq:3dV} is formally unchanged from the classical
case. The resulting structure is a cyclic differential braided
(graded) Lie algebra whose non-trivial $2$-brackets are given by
\begin{align}
\ell_{2}^{\star}\big((\xi_{1},\rho_{1})\,,\,(\xi_{2},\rho_{2})\big)&=\big([\xi_{1},\xi_{2}]_\star\,,\,-[\rho_{1},\rho_{2}]_\star+\LL_{\xi_1}^{\star}(\rho_{2})
- \LL_{\bar{\sfR}^k\xi_{2}}^\star(\bar{\sfR}_{k}\rho_{1})\big) \ , \nn
\\[4pt]
\ell_{2}^{\star}\big((\xi,\rho)\,,\,(e,\omega)\big)&=\big(-\rho \star e
+\LL_{\xi}^\star e \,,\, -[\rho,\omega]_\star+\LL_{\xi}^\star\om\big) \
, \nn \\[4pt]
\ell_{2}^{\star}\big((\xi,\rho) \,,\, (E,{\mit\Omega})\big)&=\big(-
\rho \star E+\LL_{\xi}^\star E \,,\, -\rho \star
{\mit\Omega}+\LL_{\xi}^\star{\mit\Omega} \big) \ , \nn \\[4pt]
\ell_{2}^{\star}\big((\xi,\rho)\,,\,(\CX,\CP)\big)&= 
                                                \Big(\dd x^\mu\otimes\Tr\big(\iota_{\mu}\dd
                                                (\,\bar{\sff}^{k} \rho)
                                                \dwedge \bar{\sff}_{k}
                                                \CP\big)+\LL_{\xi}^\star\CX\,,\,-\rho
                                                \star \CP
                                                +\LL_{\xi}^\star
                                                \CP\Big) \, \label{eq:braided3dbrackets} \\[4pt] 
\ell_{2}^{\star}\big((e_{1},\omega_{1})\,,\,(e_{2},\omega_{2})\big)&=-\big(\,[\omega_{1}, \omega_{2}]_{\star}
\,,\, \omega_{1}\wedge_\star
e_{2} +\bar{\sfR}^k\omega_{2} \wedge_\star \bar{\sfR}_k e_{1} \big) \
                                                                     , \nn
  \\[4pt]
\ell_{2}^{\star}\big((e,\om)\,,\,(E,{\mit\Omega})\big)&=\Big(\dd x^{\mu}\otimes \Tr  \big( \iota_\mu\dd (\,\bar{\sff}^{k}e) \dwedge \bar{\sff}_{k}E + \iota_\mu\dd(\,\bar{\sff}^k \om) \dwedge \bar{\sff}_{k}{\mit\Omega} \nn   -\iota_\mu(\,\bar{\sff}^k e) \dwedge  \dd (\,\bar{\sff}_{k} E)  \nn \\ 
&\hspace{3cm} - \iota_\mu(\,\bar{\sff}^{k} \om )\dwedge \dd
  (\,\bar{\sff}_k {\mit\Omega}) \big)\, ,\,  \bar{\sfR}^k
  E\wedge_\star \bar{\sfR}_k e - \omega \wedge_\star {\mit\Omega}
  \Big)  \ . \nn 
\end{align}

We next show that this braided $L_\infty$-algebra organises all
symmetries and dynamics of three-dimensional noncommutative gravity. Firstly, one
verifies that the action functional \eqref{eq:NC3daction} is given as in the classical case:
\begin{align}
S_{\textrm{\tiny ECP}}^{\star}[e,\om]&= \int_M\, \Tr
\Big(e\dwedge_\star\big(\dd\omega+\tfrac12\,
                 [\omega,\omega]_\star\big)\Big) \nn \\[4pt]
&= \tfrac{1}{2}\, \big\langle(e,\om)\,,\, \ell_{1}^{\star}(e,\om)
  \big\rangle_\star -\tfrac{1}{6}\,
  \big\langle(e,\om)\,,\,\ell_{2}^{\star}\big((e,\om)\,,\,(e,\om)\big)\big\rangle_\star \ .
\end{align}
By cyclicity, the Euler-Lagrange derivatives follow from the same
expansion as classically. The same expansions also hold for the
expressions of the gauge transformations and the covariance 
of the field equations. However, braided gauge invariance of the
action functional together with cyclicity result in the braided
version of Noether's identities, due to the braided Leibniz rule. In summary:
\begin{myitemize}
\item $\underline{\text{Braided gauge transformations of fields:}}$
 \begin{align}
\delta_{(\xi, \rho)}^{\star} (e,\om) = \ell_{1}^{\star}(\xi,\rho) +
   \ell_{2}^{\star}\big((\xi,\rho)\,,\,(e,\om)\big) \ .
\end{align}
\item $\underline{\text{Euler-Lagrange derivatives:}}$
\begin{align}
F^\star_{(e,\om)}=(F^\star_{e},F^\star_{\om})= \ell_{1}^{\star}(e,\om)
  -\tfrac{1}{2}\,\ell_{2}^{\star}\big((e,\om)\,,\,(e,\om)\big) \ .
\end{align}
\item $\underline{\text{Braided covariance of field equations:}}$
\begin{align}
\delta_{(\xi,\rho)}^{\star} F_{(e,\om)}^*=
  \big(-[\rho,F_e^\star]_\star+\LL_\xi^\star F_e^\star\,,\,-\rho\star
  F_\om^\star+\LL_\xi^\star F_\om^\star\big) =
  \ell_{2}^{\star}\big((\xi,\rho)\,,\,F_{(e,\om)}^\star\big) \ .
\end{align}
\item $\underline{\text{Gauge redundancy via braided Noether
      identities:}}$
\begin{align}
\ell_{1}^{\star} (F_{(e,\om)}^\star) - \tfrac{1}{2}\,\big(
  \ell_{2}^{\star}((e,\om),F_{(e,\om)}^\star)
  -\ell_{2}^{\star}(F_{(e,\om)}^\star,(e,\om))\big)+\tfrac{1}{4}\,\ell_{2}^{\star}\big(&\bar{\sfR}^{k}(e,\om)\,,\,\ell_{2}^{\star}(\bar{\sfR}_{k}(e,\om),(e,\om))\big)
                                                                                         \nn \\[4pt]
  &= (0,0) \ , \label{eq:braidedNoether}
\end{align}
where $\bar{\sfR}^{k}(e,\om):=(\bar{\sfR}^{k}e,\bar{\sfR}^{k}\om)$.
\end{myitemize}

Note the compact form that the braided
Noether identities take in this braided $L_{\infty}$-algebra
formulation: Both identities corresponding to braided local Lorentz transformations and braided
diffeomorphisms are included in \eqref{eq:braidedNoether} in a single
line. In contrast, the explicit form of the identities becomes rather
cumbersome when written out explicitly~\cite{Inprep}. For example,
the Noether identity corresponding to braided local Lorentz transformations is 
\begin{align}
\dd F^\star_{\om} - \tfrac{1}{2}\,\big(&\bar{\sfR}^k F^\star_{e} \wedge_{\star} \bar{\sfR}_{k}e - \om \wedge_{\star} F^\star_{\om} + F^\star_{e} \wedge_{\star} e - \bar{\sfR}^{k} \om \wedge_{\star} \bar{\sfR}_k F^\star_{\om}\big) \nn \\
&+\tfrac{1}{4}\, \big( -[\om,\bar{\sfR}^{k}\om]_{\star} \wedge_\star
  \bar{\sfR}_k e +\bar{\sfR}^k \om \wedge_\star \bar{\sfR}_k\om
  \wedge_\star e + \bar{\sfR}^k (\om \wedge_\star \om) \wedge_\star
  \bar{\sfR}_k e\big) = 0 \ . 
\end{align}
One may also verify the vanishing of the braided Noether identities identically (off-shell) by the braided homotopy identities.

This discussion extends to higher dimensions $d\geq4$ in an analogous
fashion, including now higher brackets of the braided
$L_\infty$-algebra where appropriate \cite{Inprep}.

\subsection{Outlook}

There are plenty of further natural questions and applications to
explore from various physical perspectives. From the perspective of noncommutative
gravity, the relation of our braided theory to the standard metric formulation of the theory~\cite{NCmetricgrav}, in which the torsion-free
condition is assumed from the outset, can be explored. From the perspective of
noncommutative field theories and braided $L_{\infty}$-algebras one
should explore the necessary and sufficient conditions, both physical
and mathematical, for the procedure we have discussed to succeed for other field
theories. For example, it is straightforward to construct a braided
noncommutative version of Yang-Mills theory~\cite{Barnes:2016cjm} and of Chern-Simons theory~\cite{Inprep}. Furthermore, it
would be interesting to explore the possible dual incarnation of these
structures as a ``braided BV--BRST'' formalism, and also the
quantization of such braided gauge field theories in comparison to
older work on braided quantum field theory
\cite{BraidedQFT,Sasai:2007me}. Lastly, our initial motivation ---
nonassociative gravity --- may be now further advanced, which in this line of
work will require yet a further generalization of our braided
$L_\infty$-algebras using the formalism of cochain twist deformations
and quasi-Hopf algebras.

From a mathematical perspective, the definition we presented of a braided
$L_\infty$-algebra has an obvious generalization as an
$L_\infty$-algebra object in any symmetric monoidal category. Given the
vast and fruitful applications of $L_\infty$-algebras to various
problems in mathematics, particularly in deformation theory, it would
be interesting to pursue the applications of these braided
generalizations. Specifically, they should play a role in operadic
constructions where the role of the symmetric group is replaced with
the braid group.

\paragraph{Acknowledgments.}
We thank the organisors of the Corfu Summer Institute
2019 for the stimulating meeting and the opportunity to present the
preliminary results of our work.
The work of {\sc M.D.C.} and {\sc V.R.} is supported by Project
ON171031 of the Serbian Ministry of Education, Science and
Technological Development. The work of {\sc G.G.} is supported by a
Doctoral Training Grant from the UK Engineering and Physical Sciences
Research Council. The work of {\sc R.J.S.} was supported by
the Consolidated Grant ST/P000363/1 
from the UK Science and Technology Facilities Council.

\end{document}